\documentclass[%
 reprint,
 amsmath,amssymb,
 aps,
]{revtex4-1}

\usepackage{graphicx}
\usepackage{dcolumn}
\usepackage{bm}

\newcommand{\beq}{\begin{equation}}
\newcommand{\eeq}{\end{equation}}
\newcommand{\beqa}{\begin{eqnarray}}
\newcommand{\eeqa}{\end{eqnarray}}
\newcommand{\beqan}{\begin{eqnarray*}}
\newcommand{\eeqan}{\end{eqnarray*}}
\newcommand{\bra}[1]{\langle{#1}|}

\newcommand{\ket}[1]{|{#1}\rangle}

\newcommand{\ip}[1]{\langle{#1}\rangle}

\newcommand{\ii}{{\mathrm{i}}}

\begin{document}

%
%
%
%
%
%
%
\noindent
\textbf{Comment on ``Catalytic coherence''}
\begin{abstract}
A comment on the Letter by J. \AA{}berg, \textit{Phys. Rev. Lett.} {\bf 113}, 150402 (2014) 
\end{abstract}

\AA{}berg has claimed in a recent Letter \cite{Aberg} that the coherence of a reservoir can be used repeatedly to perform coherent operations without ever diminishing in power to do so. The claim has particular relevance for quantum thermodynamics because, as shown in \cite{Aberg}, latent energy that is locked by coherence may be extractable without incurring any additional cost. We show here, however, that repeated use of the reservoir gives an overall coherent operation of diminished accuracy and is 
necessarily accompanied by an increased thermodynamic cost.  

We focus on an example where an energy reservoir $E$, initially in the state  $\ket{\eta_{L,l_0}(\theta)}=\sum_{l=0}^{L-1} e^{\ii l\theta}\ket{l_0+l}/\sqrt{L}$ where $\theta$ represents a phase angle and $\ket{n}$ are eigenstates of energy with eigenvalue $ns$, is used to convert a system $S$ with definite energy into a superposition of different energies \cite{phase}. Taking the claim on its face value would suggest that, provided $L$ is sufficiently large, the reservoir $E$ can be used repeatedly to transform systems in the state $\ket{\psi_0}$ to $\ket{\psi(\theta)}=(\ket{\psi_0}+e^{\ii\theta}\ket{\psi_1})/\sqrt{2}$, where $\ket{\psi_n}$ are energy eigenstates with energy $ns$, and thus it can convert a collection of $k$ systems in $\ket{\psi_0}^{\otimes k}$ to $\ket{\Psi(\theta)}=\ket{\psi(\theta)}^{\otimes k}$.  Note that  $|\ip{\Psi(\theta)|\Psi(\phi)}|=|\ip{\psi(\theta)|\psi(\phi)}|^k$ can be less than  $|\ip{\eta_{L,l_0}(\theta)|\eta_{L,l_0}(\phi)}|$ for sufficiently large values of $k$.  This would imply that the collection of systems can be used to discriminate between different values of the reservoir phase with an accuracy that is better than that of the initial reservoir state \cite{mixed state,Chefles}. As the linearity of quantum mechanics forbids this result, some qualification of the claim is clearly needed. 

What the claim and the states $\ket{\psi(\theta)}^{\otimes k}$ \cite{mixed state} leave unaccounted are any reservoir-induced correlations between the systems.  When due account is taken of such collective effects, unavoidable limitations to what can be done using the reservoir become evident. 

For example, the reservoir state $\ket{\eta_{L,l_0}(\theta)}$ is asymmetric with respect to the symmetry group $G=\{T_\phi\}$ where $T_\phi=\exp(\ii N\phi)$, $N$ is the number operator, and $\phi$ is a phase angle. A natural measure of the degree of asymmetry is given by the quantity $A_G(\cdot)$ \cite{VAWJ}, which here is $A_G(\sigma)=\ln L$ for $\sigma=\ket{\eta_{L,l_0}(\theta)}\bra{\eta_{L,l_0}(\theta)}$.  The initial state of the collection of systems $\rho=(\ket{\psi_0}\bra{\psi_0})^{\otimes k}$ is symmetric with respect to $G$, i.e. $A_G(\rho)=0$, and the asymmetry of the systems-plus-reservoir combination is just that of the reservoir, i.e. $A_G(\rho\otimes\sigma)=A_G(\sigma)$, where $G$ operates globally on composite systems, i.e. $T^{(SE)}_\phi=T^{(S)}_\phi\otimes T^{(E)}_\phi$. The asymmetry measure is non-increasing under both $G$-invariant operations ${\cal O}$, i.e. $A_G[{\cal O}(\rho\otimes\sigma)]\le A_G(\rho\otimes\sigma)$, and the partial trace, i.e. $A_G(\rho')\le A_G[{\cal O}(\rho\otimes\sigma)]$ where $\rho'={\rm Tr}_E {\cal O}(\rho\otimes\sigma)$ is the reduced state of the collection of systems \cite{VAWJ}. As the operation $V(U)$ defined in Ref. \cite{Aberg} is $G$-invariant, the effect of its repeated application between the reservoir and individual systems is bounded by the result that $A_G(\rho')\le\ln L$. Thus, the collection of systems cannot be given greater asymmetry than the initial asymmetry of the reservoir.  (For comparison, the asymmetry of the state $\ket{\psi(\theta)}^{\otimes k}$ diverges as $k$ increases \cite{mixed state}.)

To see the impact of this limitation, consider the collective fidelity $F_k=\bra{\psi(0)}^{\otimes k}\rho'\ket{\psi(0)}^{\otimes k}$ between the desired state $\ket{\psi(0)}^{\otimes k}$ and the reduced state $\rho'$  of $k$ systems generated from $\ket{\psi_0}^{\otimes k}$ using the reservoir in state $\sigma=\ket{\eta_{L,l_0}(0)}\bra{\eta_{L,l_0}(0)}$ and $k$ applications of the operator $V(U)$ where $\bra{\psi_n}U\ket{\psi_{n'}}=1/\sqrt{2}$ for $n,n'=0,1$. We find $F_k={\rm Tr}_E [(1+\Delta^{-1})^k\sigma(1+\Delta^{-1})^k/4^k]=1-k/L$, where $\Delta$ is defined in Ref. \cite{Aberg}, which reduces linearly with $k$, the number of times the reservoir is used.  This represents a serious limitation to the power of the reservoir to perform multiple coherent operations, and hence the extent to which it is truly catalytic in nature.

Moreover, the entropy of the reservoir increases each time it is used to perform a coherent operation.  In order to return the reservoir to its initial state the accumulated entropy needs to be erased.  According to Landauer's erasure principle \cite{Landauer,VB,BV}, its erasure will incur a finite physical cost.  If the coherent operations are used in a work extraction process, then any such erasure cost needs to be accounted against the net output. 

In summary, the pertinent issues when considering repeated use of the reservoir is the \textit{overall accuracy} of multiple coherent operations and any \textit{costs} associated with returning the reservoir to its initial state; we have shown the accuracy diminishes linearly with the number of operations and each operation incurs a thermodynamic cost.  Any application of the results of Ref.~\cite{Aberg} would be incomplete without paying due attention to these issues.  

\noindent
S. Bedkihal and J.A. Vaccaro\\
\hangindent=\parindent 
Centre for Quantum Dynamics, Griffith University, 
Nathan 4111, Australia

\noindent
S.M. Barnett\\
\hangindent=\parindent 
School of Physics and Astronomy, 
University of Glasgow, 
Glasgow G12 8QQ, United Kingdom


\end{document}